\documentclass[reprint,longbibliography,twocolumn,showpacs,floatfix,aps,prb]{revtex4-2}

\usepackage{amsfonts}
\usepackage[hidelinks]{hyperref}
\usepackage{graphicx}
\usepackage{amsmath}
\usepackage{amssymb}
\usepackage{physics}

\newcommand*{\Hhat}{\skew{3.5}{\hat}{H}}

\begin{document}
\title{Zigzag antiferromagnets in the SU(3) Hubbard model on the square lattice}

\author{Stijn V. \surname{Kleijweg}}
\email{s.v.kleijweg@uva.nl}

\author{Philippe \surname{Corboz}} 

\affiliation{Institute for Theoretical Physics Amsterdam, University of Amsterdam, Science Park 904, 1098 XH Amsterdam, The Netherlands}

\date{June 17, 2025}

\begin{abstract}
SU($N$) Hubbard models exhibit a rich variety of phases, which may be realized through quantum simulation with ultracold atomic gases in optical lattices. In this work we study the Mott insulating phases of the SU(3) Hubbard model at 1/3-filling using infinite projected entangled-pair states, optimized with both imaginary time evolution and variational optimization. In the limit of strong interactions we reproduce the antiferromagnetic 3-sublattice ordered state previously identified in the SU(3) Heisenberg model. At intermediate interaction strength we find antiferromagnetic states
exhibiting zigzag patterns of different lengths, in agreement with previous Hartree-Fock and constrained-path auxiliary-field quantum Monte Carlo calculations. We study the color order parameter and energy anisotropy, which are discontinuous across the phase transitions. Finally, we analyze the different energy contributions in two competing phases, identifying low-energy bonds at the corners of the zigzag that help stabilize the zigzag states.
\end{abstract}

\maketitle

\section{Introduction}
SU($N$) (Fermi-)Hubbard models constitute a generalization of the famous Hubbard model~\cite{hubbard1963,arovas2022,qin2022} to $N > 2$ different species of fermions. To high precision, these models are realized in ultracold gases of alkaline-earth(-like) atoms in optical lattices~\cite{wu2003,honerkamp2004,rapp2008,cazalilla2009,gorshkov2010,cazalilla2014,ibarra2024,footnote_alkali}. Such analog quantum simulation of SU($N$) Hubbard models has been achieved using strontium and ytterbium atoms~\cite{taie2012,scazza2014,zhang2014,hofrichter2016,ozawa2018,taie2022,tusi2022,pasqualetti2024}. The spin (or flavor) correlations can be measured using quantum gas microscopes, which have been used successfully for SU(2) fermions~\cite{bakr2009,sherson2010,bloch2012,cheuk2015,edge2015,haller2015,omran2015,parsons2015,parsons2016, gross2017,mazurenko2017,hartke2020,ji2021,koepsell2021,gross2021}. With ongoing progress in SU($N$) microscopes~\cite{miranda2015,yamamoto2016,miranda2017,okuno2020,schafer2020,taie2022,buob2024,mongkolkiattichai2025} a comparison of the experimental magnetic correlations to the theoretical and numerical predictions may become possible in the future.

The SU($N$) Hubbard models are also inherently interesting from a theoretical perspective, as are their effective SU($N$) Heisenberg models in the strongly interacting limit and at integer filling. On two-dimensional lattices, numerous intriguing phases have been discovered in these models, including various color-ordered states, generalized valence bond solids, and quantum spin liquids~\cite{affleck1988,marston1989,read1989,read1990,harada2003,assaad2005,hermele2009,toth2010,corboz2011a,hermele2011,bauer2012,corboz2012a,corboz2012b,corboz2013,nataf2014,chen2016,dufour2016,nataf2016a,nataf2016b,nie2017,hafez-torbati2018,hafez-torbati2019,hafez-torbati2020,chung2019,liu2019,feng2023,ibarra2023,botzung2024a,botzung2024b,chen2024,schlomer2024,zhang2025} (see also Table 1 in~\cite{ibarra2024} for an overview, including finite temperature results). The realization of these phases depends on the parameters of the model, including the value of $N$, interaction strength, lattice geometry, particle density~\cite{schlomer2024,bohler2025}, flavor-imbalance~\cite{zhang2025} and temperature. Taking all these parameters together leads to a considerable space of models that remains largely unexplored.

\begin{figure}[t]
	\centering
	\includegraphics[width=0.99\linewidth]{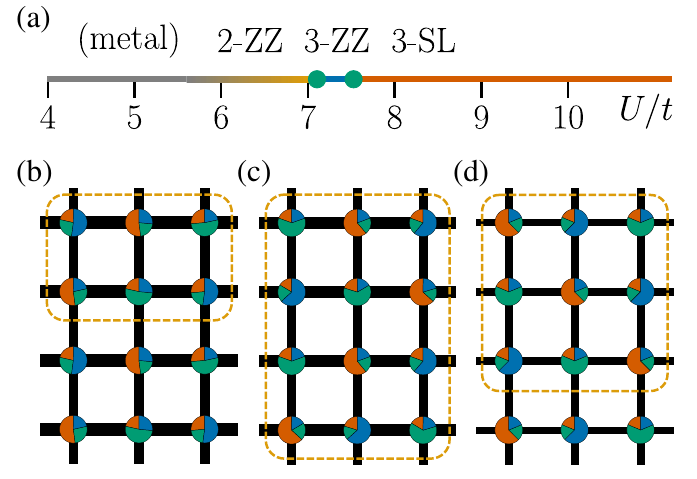}
	\caption{(a)~The iPEPS ground state phase diagram of the square lattice SU(3) Hubbard model at particle density $n = 1$, based on bond dimension $D=24$ simple-update results. The three ordered phases are (b)~the 2-zigzag, (c)~the 3-zigzag and (d)~the 3-sublattice phase. The rectangular unit cells are indicated by the orange boxes, the local color densities by the pie charts. The bond width scales with the bond energy (note the anisotropy in the two finite-length zigzags). The phase transitions between the ordered phases occur around $U/t = 7.1$ and $U/t = 7.5$. The metal-insulator transition was found to occur in the range $U/t=5.5 \text{--} 6$ in Refs.~\cite{feng2023,pasqualetti2024}.}
	\label{fig:pd}
\end{figure}

In this work we study the SU(3) Hubbard model on the square lattice at 1/3-filling (one particle per site), focusing on the Mott insulating phases at zero temperature. We use infinite projected entangled-pair states (iPEPS), a tensor network ansatz that represents two-dimensional states in the thermodynamic limit, which has been applied successfully to many strongly correlated systems, see e.g.\ Refs.~\cite{corboz2011a,bauer2012,corboz2012a,corboz2013,corboz2014a,corboz2014b,nataf2016a,liao2017,niesen2017,chen2018,jahromi2018,lee2018,niesen2018,yamaguchi2018,chung2019,ponsioen2019,gauthe2020,kshetrimayum2020,lee2020,czarnik2021,hasik2021,jimenez2021,gauthe2022,hasik2022,liu2022,peschke2022,shi2022,sinha2022,ponsioen2023,xu2023,hasik2024,schmoll2024,weerda2024}. In the limit of strong interactions, we identify a 3-sublattice diagonal stripe phase, consistent with previous results for the SU(3) Heisenberg model~\cite{toth2010,bauer2012}. Decreasing the interaction strength, we find two more ordered states which follow a zigzag pattern, similar to the 3-sublattice state, but with turns in the stripes. The first zigzag has a length of 3 sites along the diagonal between the corners, and the second zigzag has a length of 2, as summarized in the phase diagram in Fig.~\ref{fig:pd}. These are the same phases as those found in a recent Hartree-Fock and constrained-path auxiliary-field quantum Monte Carlo (CP-AFQMC) study~\cite{feng2023}.

\section{Model}
The SU($N$) Hubbard Hamiltonian is given by
\begin{equation}
	\Hhat = -t \sum\limits_{\langle i,j \rangle, \, \alpha} \left( \skew{3}{\hat}{c}^\dagger_{i,\alpha} \skew{3}{\hat}{c}_{j,\alpha} + \mathrm{H.c.} \right)  + U \sum\limits_{i, \, \alpha<\beta} \skew{1}{\hat}{n}_{i,\alpha} \skew{1}{\hat}{n}_{i,\beta},
\end{equation}
where first sum runs over the square lattice nearest-neighbor sites $i$ and $j$, and the second sum over the three SU(3) flavors (or colors) $\alpha \in \{1,2,3\}$. The operator $\skew{3}{\hat}{c}_{i,\alpha}$ annihilates a fermion of flavor $\alpha$ on site $i$, and the number operator is defined as $\skew{1}{\hat}{n}_{i,\alpha} \equiv  \skew{3}{\hat}{c}^\dagger_{i,\alpha} \skew{3}{\hat}{c}_{i,\alpha}$. 
$U$ is the strength of the on-site interaction, and $t$ is the tunneling amplitude ($t=1$). The particle density is fixed to $n = 1$.

\section{Methods}
iPEPS is a two-dimensional variational tensor network ansatz for ground states in the thermodynamic limit~\cite{verstraete2004,jordan2008}. A unit cell of one or more tensors is repeated infinitely, where the size of the unit cell is chosen to be compatible with the translational symmetry breaking~\cite{jordan2008,corboz2011b}. iPEPS belongs to the class of tensor network algorithms, like the well-known density matrix renormalization group~\cite{white1992}, which do not suffer from the negative-sign problem~\cite{troyer2005}. The tensors that constitute the iPEPS have a physical index with dimension equal to the local Hilbert space, $2^3=8$ in this case. The remaining indices (bonds) have dimension $D$, which controls the number of variational parameters and thus the accuracy of the ansatz. Fermionic systems can be treated at virtually no extra cost by incorporating the fermionic parity symmetry and tracking crossing tensor legs~\cite{corboz2009,corboz2010} (other formalisms exist as well, see~\cite{barthel2009,shi2009,kraus2010,mortier2025}). Other symmetries can also be built into the ansatz, such as global Abelian symmetry~\cite{bauer2011,singh2011}. Here we impose $U(1) \times U(1) \times U(1)$ symmetry, corresponding to particle conservation per flavor. Implementing these symmetries leads to a block sparse structure of the tensors, which reduces the computational cost, allowing us to reach higher bond dimensions.

Unlike the matrix product state, the iPEPS cannot be contracted exactly. We use the corner transfer matrix renormalization group (CTMRG)~\cite{nishino1996,orus2009,corboz2014b} to do an approximate contraction, where the infinite system surrounding a center site is represented by an environment which consists of four corner and four edge tensors. The accuracy of the contraction is controlled by the environment bond dimension, labeled by $\chi$. In practice, we use a sufficiently large value of $\chi$ such that the contraction error is small compared to the symbol sizes in the plots.

Two classes of methods exist for the optimization of the ansatz: imaginary time evolution and variational optimization. In the former the ansatz is projected onto the ground state by repeatedly applying $\mathrm{e}^{-\tau \Hhat}$ (Trotter-Suzuki decomposed into two-site operators). We employ the simple update (SU)~\cite{jiang2008} and the (fast) full update~\cite{jordan2008,orus2009,phien2015} to perform the involved truncations. Variational optimization methods are based on a direct energy minimization with respect to the tensor elements, which is more accurate, but more computationally expensive~\cite{corboz2016,vanderstraeten2016}. We use automatic differentiation (AD)~\cite{liao2019} to compute the required gradients. The complementary approach allows us to obtain results at high bond dimension $D$ with simple update, to be cross-checked with refined data at smaller $D$ from AD. Typically the AD calculations are initialized from SU states at the same $D$, sometimes with an additional full-update optimization before continuing with AD.

\section{Results}
In this section we first describe the results at large $U/t$, where the 3-sublattice diagonal stripe phase we find is consistent with previous results in the Heisenberg limit. Then we consider intermediate $U/t$, where two more ordered states stabilize: antiferromagnets arranged in zigzag patterns of different lengths. We analyze the competition between these states as a function of the bond dimension $D$ at different points in the phase diagram, followed by a discussion of the color order and energy anisotropy across the phase transitions. Finally, we investigate the competition between the 3-sublattice state and the intermediate zigzag state in more detail, splitting the total energy into different contributions, and comparing the real-space structure of these.

\subsection{Large $U/t$}

\begin{figure}[b]
	\centering
	\includegraphics[width=\linewidth]{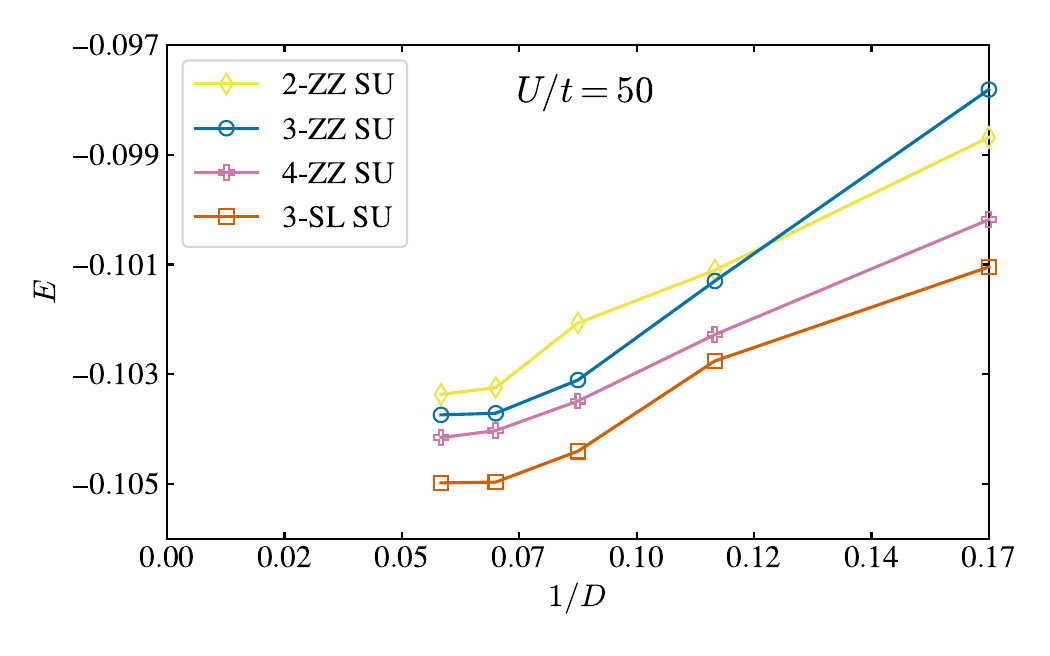}
	\caption{Simple-update results for the energy per site against the inverse bond dimensions up to $D = 18$, in the limit of large $U/t = 50$. There is a clear separation in energy between the 3-sublattice state and the zigzags of lengths 2, 3 and 4.}
	\label{fig:heisenberg}
\end{figure}

Focusing first on the limit of strong interactions at $U/t = 50$, we obtain a 3-sublattice antiferromagnet with diagonal stripes as the ground state. This is the same state as found in previous studies of the corresponding Heisenberg model~\cite{toth2010,bauer2012}. The state is depicted in Fig.~\ref{fig:pd}(d), where the local color densities are plotted in pie charts on the sites. Picking a color and following its majority sites, one recognizes the diagonal stripes.

There are several states that compete with the 3-sublattice (3-SL) state, in particular antiferromagnetic states with a zigzag pattern, as illustrated in Fig.~\ref{fig:pd}(b) and (c). They can be regarded as modifications of the 3-sublattice state, in which the diagonal stripes undergo a 90-degree turn every $l$ lattice sites along the diagonal. We call $l$ the \textit{length} of the zigzag, denoting e.g.\ the state in Fig.~\ref{fig:pd}(b) the 2-zigzag (2-ZZ) state. The 3-sublattice state can be interpreted as an infinite zigzag, and all the $l\ge2$ competing states belong to a family of zigzags.

In Fig.~\ref{fig:heisenberg} the competition between the 3-sublattice state and the 2-, 3- and 4-zigzag states is shown for $U/t = 50$. We consider bond dimensions up to $D = 18$, where it becomes clear that the 3-sublattice state is stabilized as the ground state against these competing states. Another competing state is the 2-sublattice state with two colors, which we find has a higher energy (not shown), consistent with the Heisenberg case~\cite{bauer2012}.

\begin{figure}[tb]
	\centering
	\includegraphics[width=\linewidth]{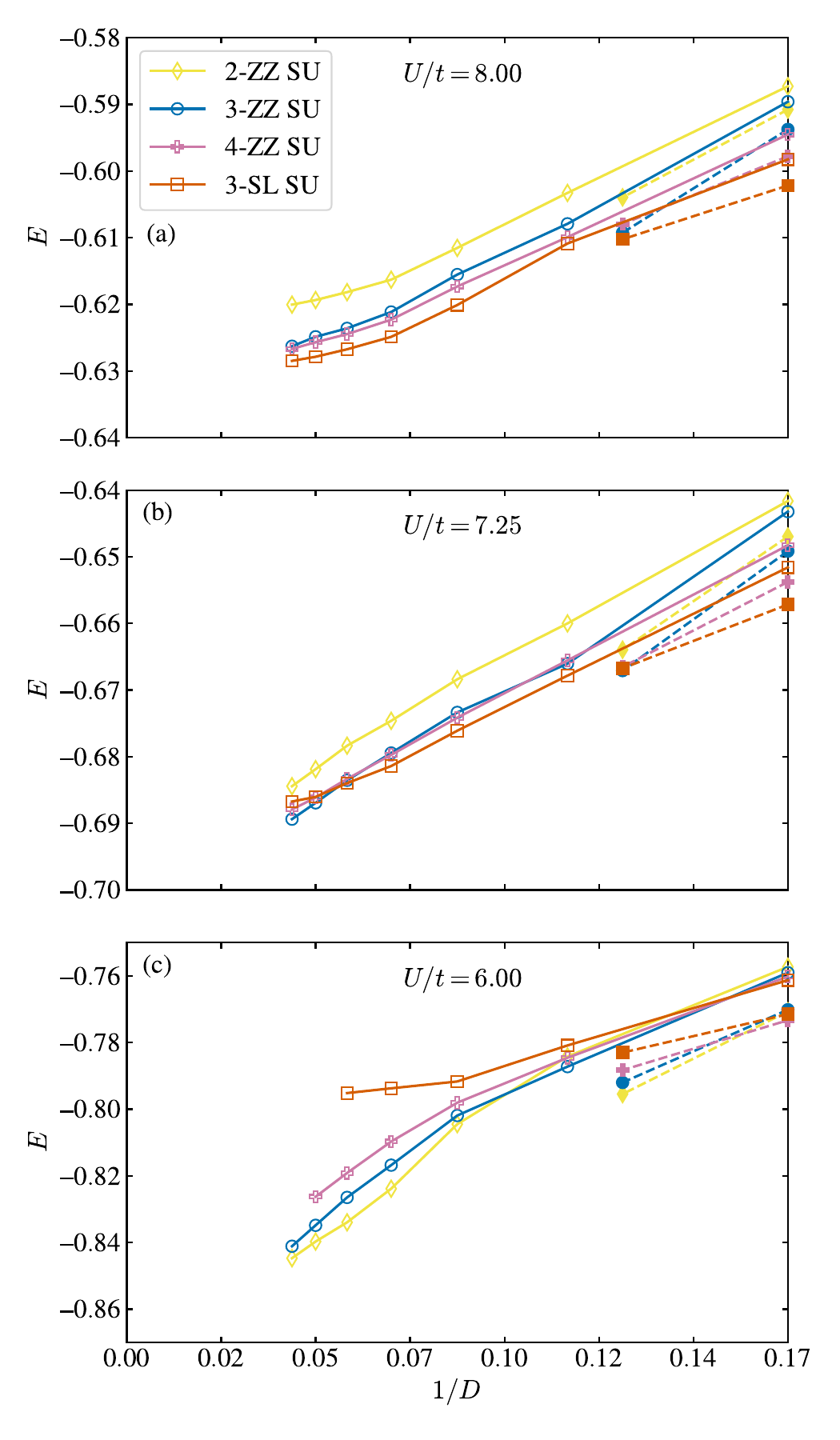}
	\caption{The energy per site against the inverse bond dimension at three values of $U/t$ where the three different phases are stabilized. The large bond dimension results up to $D = 24$ have been obtained with the simple update (open symbols). The full symbols with dashed lines are the results from automatic differentiation ($D = 6, 8$). The trends up to $D = 24$ indicate the ground states, (a)~3-sublattice, (b)~3-zigzag, (c)~2-zigzag. The 4-zigzag state is close in energy but is never the ground state.}		
	\label{fig:oneoverD}
\end{figure}

\subsection{Zigzag states at intermediate $U/t$}
The competition among the states described in the previous section changes upon decreasing $U/t$. In Fig.~\ref{fig:oneoverD} we show the energy against $1/D$ for different points in the phase diagram. In Fig.~\ref{fig:oneoverD}(a), focusing first on the large $D$ simple-update results, the 3-sublattice state is still the lowest in energy at $U/t = 8$. However, the 3-zigzag and 4-zigzag states are now much closer in energy than at $U/t = 50$ (Fig.~\ref{fig:heisenberg}). Decreasing $U/t$ further, Fig.~\ref{fig:oneoverD}(b) demonstrates the stabilization of the 3-zigzag state. The competition is close, but the calculations up to $D = 24$ indicate that the 3-zigzag state is stabilized at $U/t = 7.25$. Upon lowering $U/t$ further there is a second phase transition: the shortest 2-zigzag state is stabilized, as shown for $U/t = 6$ in Fig.~\ref{fig:oneoverD}(c).

The AD results (filled symbols in Fig.~\ref{fig:oneoverD}) reach a smaller maximum bond dimension, since they are more computationally expensive. While shifted to lower energies (as expected since the method is more accurate at a given $D$), the trends of the states are similar to the SU results, and the lowest energy states are the same, providing further confirmation of the zigzag ground states at intermediate $U/t$.

\begin{figure}[t]
	\centering
	\includegraphics[width=\linewidth]{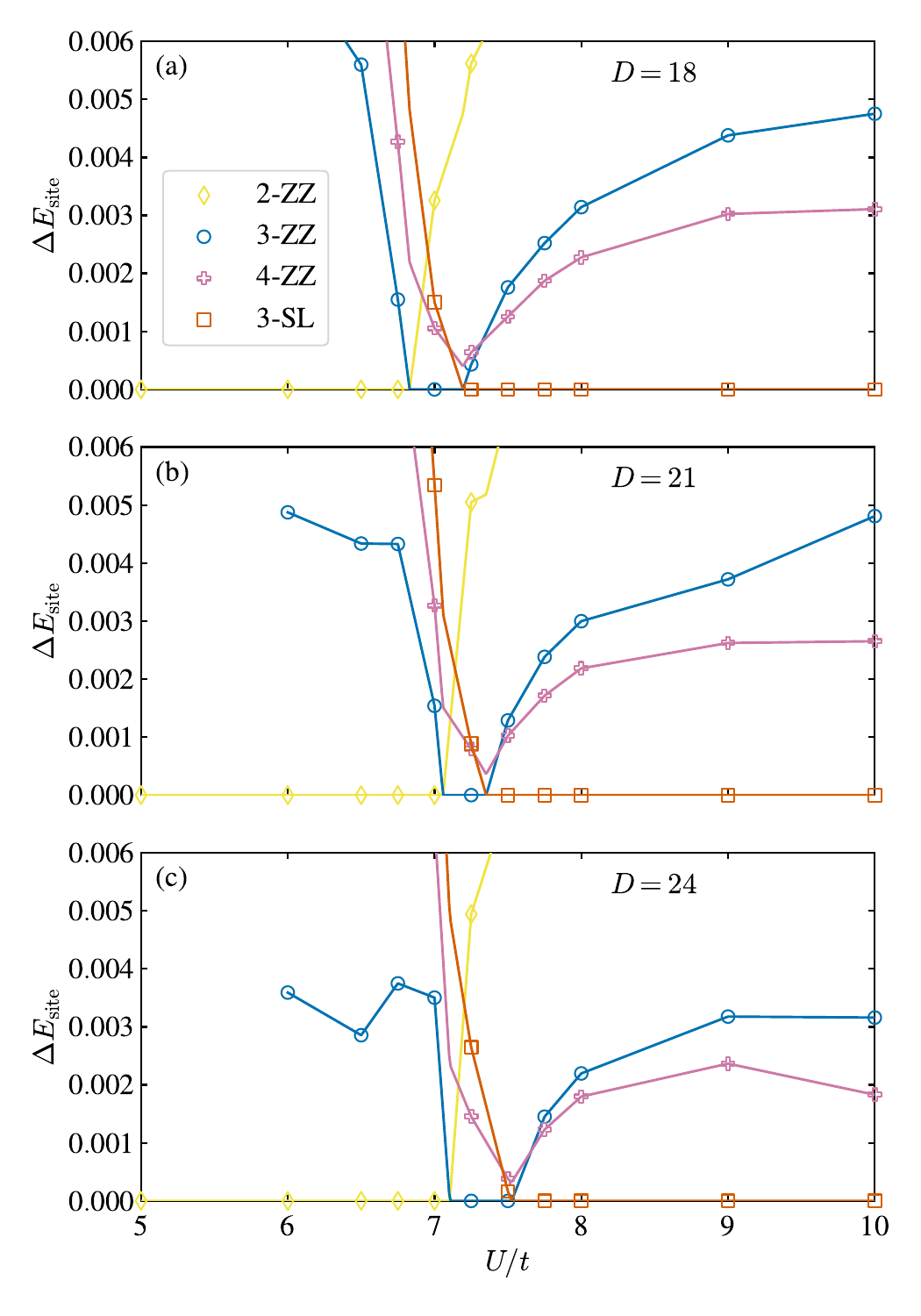}
	\caption{The energy difference between the competing states against $U/t$ for different bond dimensions, where the lowest energy is subtracted from the others (simple update). Thus, the data on the $x$-axis shows the phase diagram at the corresponding value of $D$. The 4-ZZ state is competing, but not stabilized.}
	\label{fig:pdlinint}
\end{figure}

\begin{figure}[t]
	\centering
	\includegraphics[width=\linewidth]{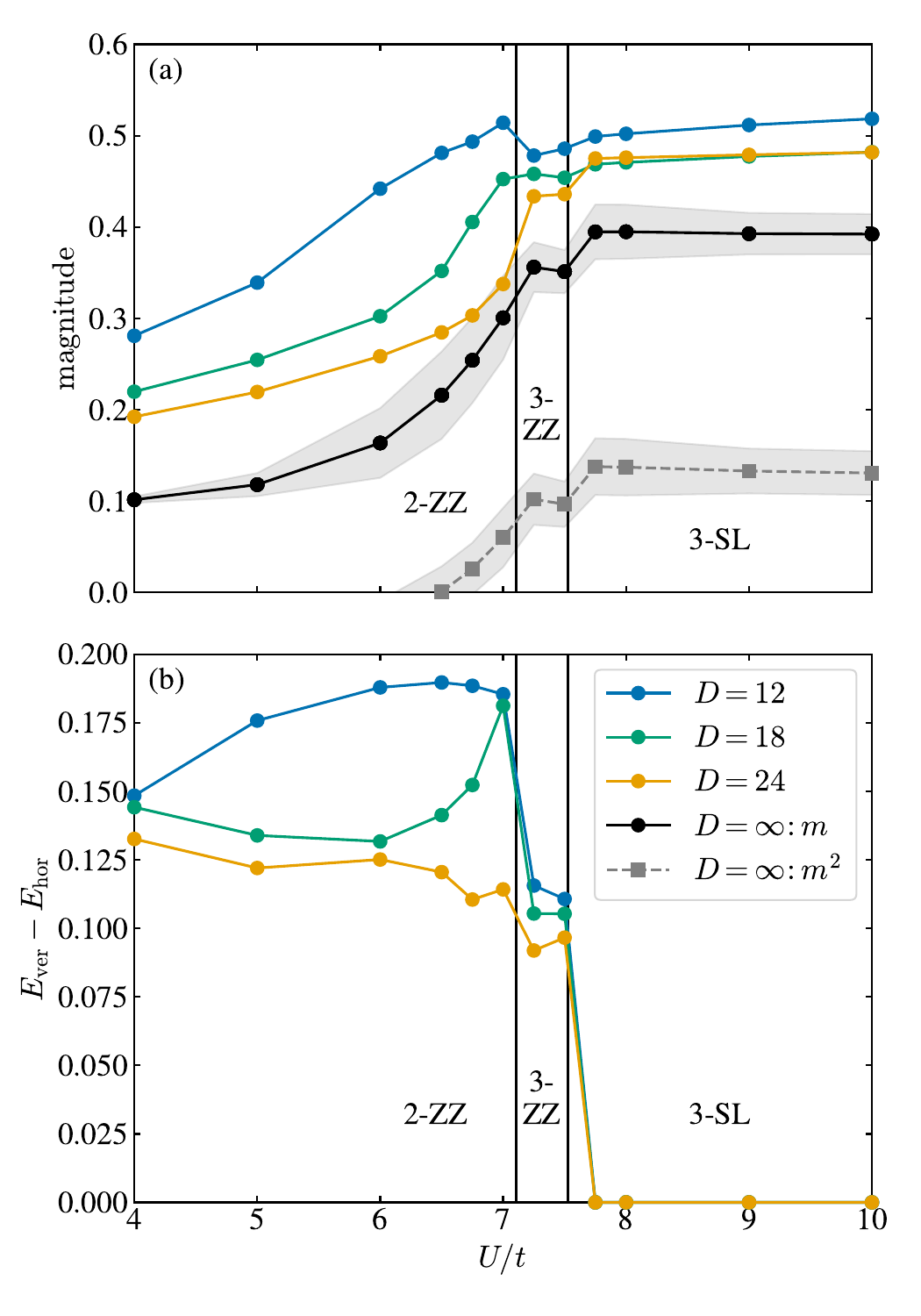}
	\caption{Order parameters as a function of $U/t$. (a)~Magnitude of the color order parameter $m$ as a function of $U/t$ for the different phases. The solid colored lines show finite bond dimension data for $D = 12, 18, 24$. The linear extrapolation of $m$ ($m^2$) is indicated by the solid (dashed) line. We have used all bond dimensions $D = 6, 9, ..., 24$ for the extrapolation, with the error bar $\sigma$ represented by the shading. (b)~Energy anisotropy as a function of $U/t$ for the same finite bond dimensions as in (a), which is zero for the 3-sublattice state. Between the ordered phases, both observables display discontinuities across the transitions.}
	\label{fig:order}
\end{figure}

A different view on the energy differences among the phases, as function of $U/t$, is shown in Fig.~\ref{fig:pdlinint}. Starting from the high-bond-dimension simple-update results, we do a linear interpolation of $E_\mathrm{site}$ in $U/t$, per state and bond dimension. Then the lowest energy is subtracted from the others at each value of $U/t$. Thus, the data on the $x$-axis shows the phase diagram.

At the largest bond dimension $D=24$, the resulting transition values are $U/t=7.5$ and $U/t=7.1$ for the first and second transition. The transition values exhibit small shifts to larger $U/t$ with increasing bond dimension, of the order of $2\%$ and $0.7\%$ between $D=21$ and $D=24$ for the two transitions. Our transition values are slightly lower than those from the CP-AFQMC study of this model: 8.3 and 7.7, but higher than the predictions from Hartree-Fock: 5.65 and 4.75~\cite{feng2023}.

The sequence of zigzags of lengths $2, 3, \infty$ in the phase diagram raises the question of whether the 4-zigzag is stabilized. The energy hierarchy of the zigzags is reversed between the large and small $U/t$ limits:
at large $U/t$ the energy of the state decreases with the length of the zigzag, and vice versa at small $U/t$. Thus, the monotonous order must reverse at some intermediate value, resulting in the small region where the 3-zigzag is stabilized. As we decrease $U/t$, zigzags with $l < \infty$ gain energy and overtake the 3-sublattice state, which explains why the 4-zigzag is so competitive around the first phase transition. However, the 4-zigzag is seemingly not stabilized at any $U/t$.

Finally, we note that we have checked for competing color-imbalanced states, in which the individual color densities deviate from 1/3. However, these states have a higher energy than the color-balanced zigzag states.

\subsection{Order parameters}
In Fig.~\ref{fig:order}(a) we show finite-bond-dimension data and extrapolations for the color order parameter, which measures the strength of the color order. It is defined as
\begin{equation}
	m = \sqrt{\frac{3}{2} \sum_{\alpha=1}^{3} \left(\expval{n_\alpha} - \frac{1}{3}\right)^2},
\end{equation}
and can be interpreted as a generalized magnetization. The above expression is averaged over the unit cell. Since the functional form of the order parameter in $D$ is unknown, we have performed a linear extrapolation in $1/D$ of both $m$ and $m^2$, based on the data from all bond dimensions ($D = 6, 9, \ldots, 24$).

The order is strongest in the 3-sublattice phase, and drops discontinuously to a slightly lower value at the transition to the 3-zigzag, suggesting a first order transition. There is another discontinuity across the second phase transition, and the order is weaker in the 2-zigzag phase, as is also visible in the pie charts in Fig.~\ref{fig:pd} (the dominant color densities are smaller).

The extrapolation of $m$ remains finite at small $U/t$. However, the square of the order parameter vanishes around $U/t = 6.5(4)$, indicating the absence of color order. This result is compatible with the metal-insulator transition in the range $U/t = 5.5 \text{--} 6$ found in the CP-AF and determinant QMC studies~\cite{feng2023,ibarra2023}. We note that, due to increasing entanglement upon approaching the non-interacting limit $U/t = 0$, it is challenging to accurately determine the metal-insulator transition with iPEPS. Nevertheless, our results suggest that the 2-zigzag phase remains stable over a finite range of $U/t$.

Fig.~\ref{fig:order}(b) shows the energy anisotropy of the different phases, defined as the difference between the average vertical and horizontal bond energies. While the 3-sublattice phase is isotropic, the zigzag phases exhibit stronger bonds in the horizontal direction (perpendicular to the zigzag, cf.\ the orientation in Fig~\ref{fig:pd}(b) and (c)). Like the color order, the energy anisotropy is discontinuous across the phase transitions between the ordered phases.

\subsection{Energy contributions}
\begin{figure}[t]
	\centering
	\includegraphics[width=\linewidth]{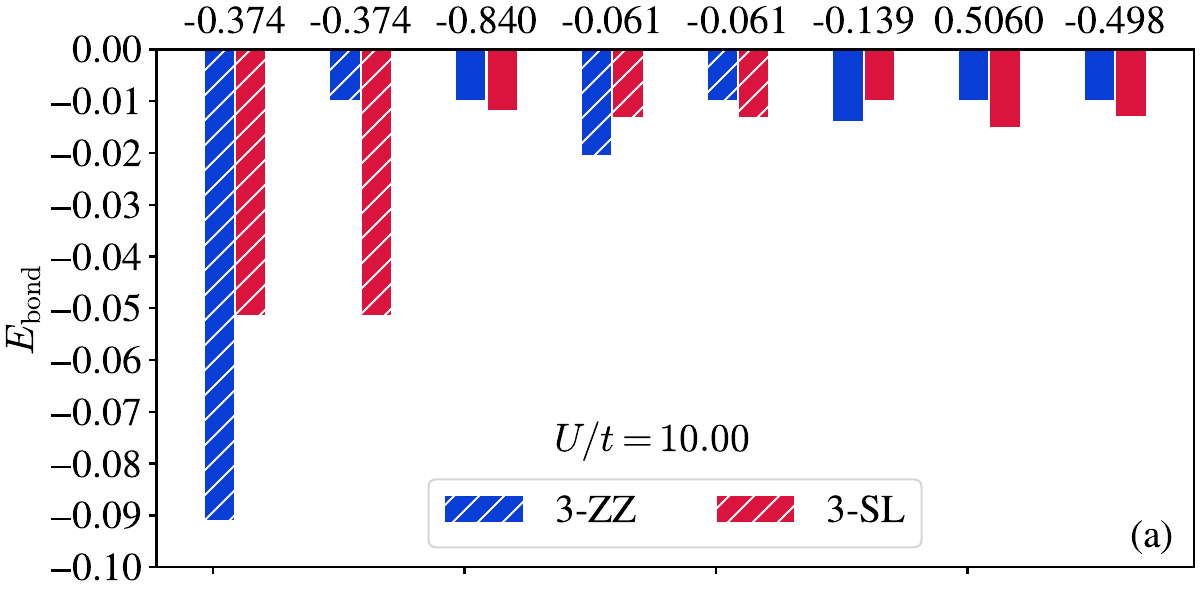}\\
	\vspace{0.2cm}
	\includegraphics[width=\linewidth]{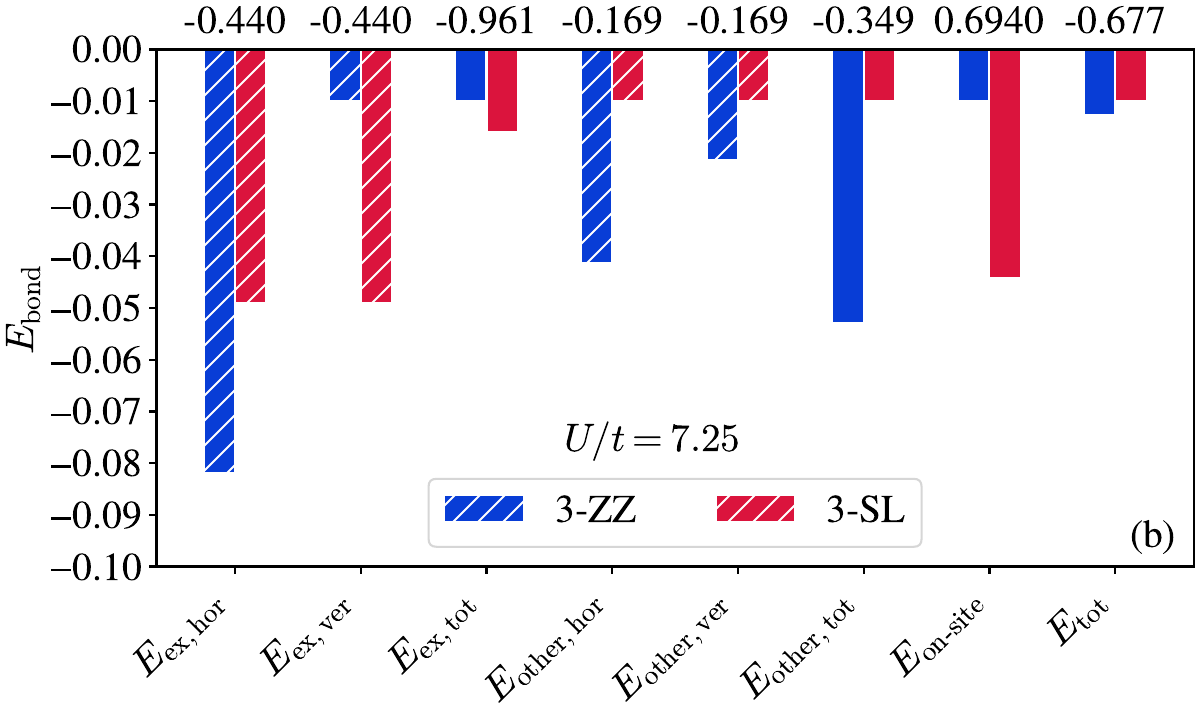}
	\caption{The energy contributions for the 3-zigzag and 3-sublattice states at (a)~$U/t = 10$ and (b)~$U/t = 7.25$. The two categories of kinetic processes are split into horizontal and vertical, indicated by shaded bars. The bars are offset from zero by the value shown above the bars to magnify the energy differences.}
	\label{fig:kincontr}
\end{figure}

To gain a better understanding of the energetics of the different phases, we analyze the various energy contributions of two of the ordered states in Fig.~\ref{fig:kincontr}. The tunneling (kinetic) term is divided into the Heisenberg superexchange energy $E_\mathrm{ex}$ (matrix elements between two singly occupied sites and a doubly occupied site and a hole) and all other processes $E_\mathrm{other}$, e.g.\ a fermion hopping from a doubly occupied site to a singly occupied site. The two categories sum to all possible hopping processes, and are further divided into horizontal and vertical contributions. Finally, with the on-site term (which is positive), these sum to the total energy per site.

We notice first in Fig.~\ref{fig:kincontr}(a) that the 3-zigzag state has a lower superexchange energy in the horizontal direction at $U/t = 10$. However, the 3-sublattice state has a lower $E_\mathrm{ex}$ in the vertical direction, and they almost compensate, with a slight energetic advantage for the 3-sublattice state. The 3-zigzag state has a lower energy for $E_\mathrm{other}$, but the 3-sublattice state has a lower on-site energy, due to a reduced density of doubly occupied sites, leading ultimately to its stabilization. Decreasing $U/t$ to 7.25 in Fig.~\ref{fig:kincontr}(b) the situation is reversed: a significant energy gain due to the other kinetic processes stands out for the 3-zigzag state, while the 3-sublattice state gains much (but less) in the on-site energy. This results in the stabilization of the 3-zigzag state.

\begin{figure}[t]
	\includegraphics[width=\linewidth]{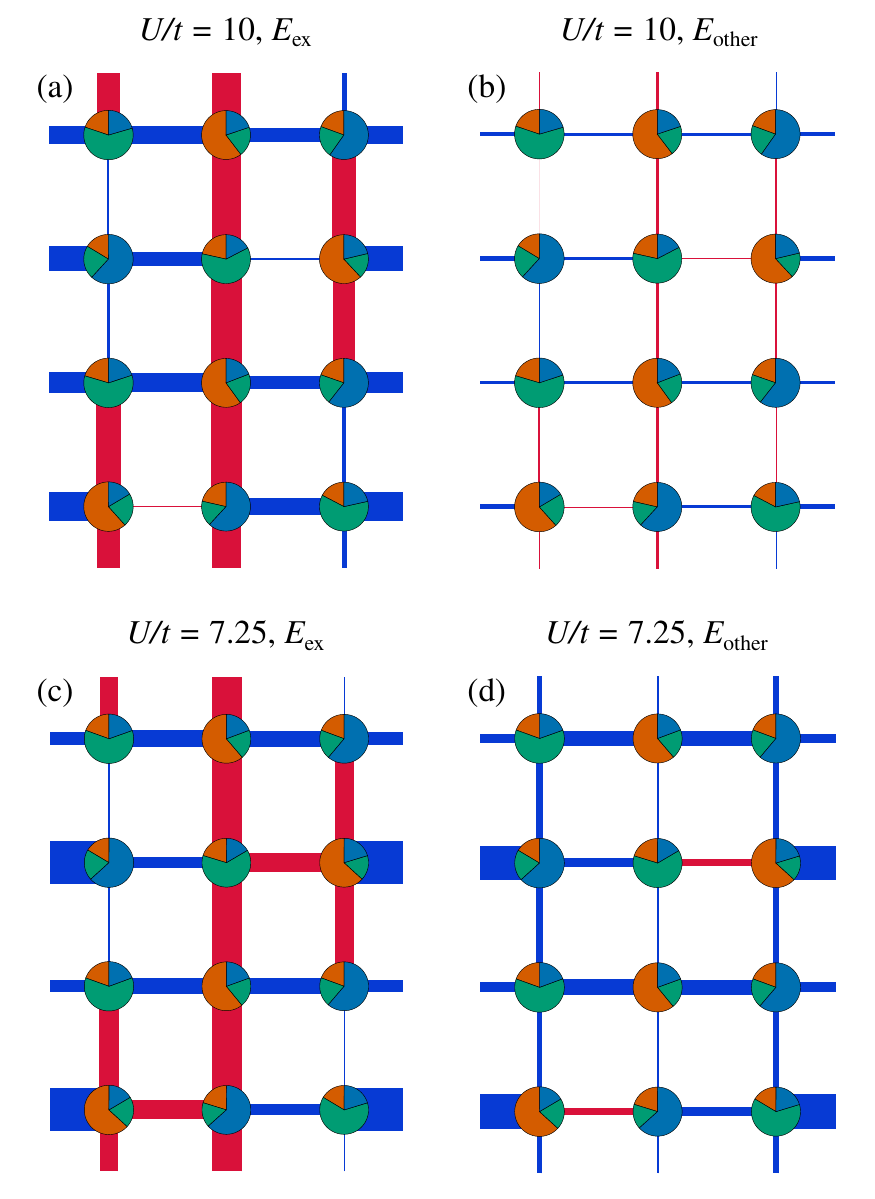}
	\caption{The energy differences between the 3-zigzag and the 3-sublattice states for a single unit cell of the 3-zigzag state, from $D = 24$ simple-update data, including only hopping contributions of the red particle. The 3-SL bond energies are taken as a reference and subtracted from the the corresponding 3-ZZ bonds with the same two dominant colors. (a), (b)~are at $U/t=10$, (c), (d)~at $U/t=7.25$. The Heisenberg superexchange energy $E_\mathrm{ex}$ is shown in the left two diagrams, the remaining kinetic processes $E_\mathrm{other}$ in the right two. Blue (red) bonds indicate a lower (higher) energy for the 3-ZZ state.}
	\label{fig:bondplots}
\end{figure}

To investigate the real space structure of this competition, the bond energies are compared in Fig.~\ref{fig:bondplots}. The diagrams (a) and (b) show the energy difference between the 3-ZZ and 3-SL states at $U/t = 10$, the diagrams (c) and (d) at $U/t = 7.25$. We consider a single color (red) and separate the energy contributions again into Heisenberg superexchange $E_\mathrm{ex}$ and other kinetic processes $E_\mathrm{other}$, shown in the left and right column of Fig.~\ref{fig:bondplots}, respectively. To compare the states we subtract the corresponding bonds, e.g.\ the bond between a majority green and a majority red site in the 3-SL state from the equivalent bond in the 3-ZZ state.

In Fig.~\ref{fig:bondplots}(a) and (c) we observe that the disadvantage in vertical superexchange energy for the 3-ZZ state stems from the vertical bonds connected to the red sites. Second, focusing on the other kinetic processes, a notable change occurs in the 3-ZZ state compared to the 3-SL state when decreasing $U/t$, as shown in Fig.~\ref{fig:bondplots}(b) and (d): there is a significant gain in $E_\mathrm{other}$, especially along the horizontal direction. Finally, the bonds extending outward from the zigzag corners are special locations that gain a significant amount of both $E_\mathrm{ex}$ and $E_\mathrm{other}$. The combined effect of these energy gains stabilizes the 3-ZZ state as the ground state.

\section{Discussion}
We have determined the phase diagram of the SU(3) Hubbard model on the square lattice at a filling of one particle per site, focusing on the insulating, color-ordered states. It is based on high-bond-dimension simple-update iPEPS results, with variational optimization at smaller bond dimensions showing similar trends of the energy. The ordered states all belong to a family of zigzag states, where the known 3-sublattice Heisenberg state can be considered an infinitely long zigzag state. Upon lowering $U/t$ the two shortest zigzag states are stabilized; first the $l = 3$ and then the $l = 2$ state. The $l = 4$ zigzag state competes closely in a narrow range of $U/t$ across the first phase transition, but it remains higher in energy. The phase transitions occur at values of $U/t$ close to those found in a CP-AFQMC study~\cite{feng2023}.

We have computed the color order parameter and the energy anisotropy, both of which are discontinuous across the phase transitions between the zigzag and 3-sublattice states. Obtaining a precise estimate of the color order parameter in the infinite $D$-limit is challenging, because the functional dependence on $D$ is not known. However, we have observed that, based on a $1/D$ extrapolation of $m^2$, the color order parameter vanishes around $U/t = 6.5$, which is compatible with the value of the metal insulator transition found in two previous studies~\cite{feng2023,ibarra2023}.

Finally, we have investigated the energetics of the 3-zigzag and 3-sublattice states, and observed that the zigzag state gains energy through the non-Heisenberg kinetic processes, resulting in a lower energy at intermediate values of $U/t$ despite a larger double occupancy (larger on-site repulsion). We have identified the bonds extending away from the zigzag corners as special locations where a particularly large energy gain occurs.

Our work offers a complementary view on the phase diagram of the square-lattice SU(3) Hubbard model, alongside previous Hartree-Fock, CP-AFQMC and determinant QMC studies~\cite{feng2023,ibarra2023}. The results obtained here, in particular the antiferromagnetic correlations, should be observable in ultracold atomic gases in optical lattices in the (near) future. This provides an exciting prospect of validating the numerical predictions of the zigzag states in experiments.

\acknowledgments
This project has received funding from the European Research Council (ERC) under the European Union's Horizon 2020 research and innovation programme (grant agreement No. 101001604).

\bibliography{refs.bib}

\end{document}